\documentstyle[sprocl]{article}

\def\be{\begin{equation}}
\def\ee{\end{equation}}
\def\bea{\begin{eqnarray}}
\def\eea{\end{eqnarray}}

\newcommand{\M}{{\cal M}}

\newcommand{\ket}[1]{\vert\,{#1}\rangle}
\renewcommand{\bar}{\overline}
\newcommand{\ie}{{\it i.e.}}
\newcommand{\eg}{{\it e.g.}}
\newcommand{\etal}{{\em et al.}}

\newcommand{\half}{{$\frac{1}{2}$}} 

\def\ru1{\rule[-0.4truecm]{0mm}{1truecm}}

\begin{document}

\title{Light-Cone Wavefunctions and the Intrinsic Structure of
Hadrons}

\author{STANLEY J. BRODSKY}

\address{Stanford Linear Accelerator Center \\
Stanford University, Stanford, California 94309 USA\\E-mail:
sjbth@slac.stanford.edu}

\maketitle\abstracts{ The light-cone Fock-state representation of QCD
encodes the properties of a hadrons in terms of frame-independent
wavefunctions.  A new
type of jet production reaction, ``self-resolving diffractive
interactions" can provide direct information on the light-cone
wavefunctions of hadrons in terms of their quark and gluon degrees of
freedom as well as the composition of nuclei in terms of their nucleon
and mesonic degrees of freedom. The relation of the intrinsic sea to the
light-cone wavefunctions is discussed.  The decomposition of light-cone
spin is illustrated for the quantum fluctuations of an electron.  }

\section{Introduction}

The most challenging nonperturbative problem in quantum chromodynamics
is the solution of the bound state problem; \ie, to determine the
spectrum and structure of hadrons in terms of their quark and gluon
degrees of freedom.  Ideally, one wants a
frame-independent, quantum-mechanical description of had\-rons at the
amplitude level capable of encoding multi-quark and gluon
momentum, helicity, and flavor correlations in the form of universal
process-independent hadron wavefunctions.  Remarkably, the light-cone Fock
expansion allows just such a unifying representation.

In the light-cone Hamiltonian method, QCD is quantized at a fixed
light-cone time $\tau = t +z/c$.  \cite{Dirac:1949cp} The generator of
light-cone time translations $P^- =P^0 - P_z = i {\partial \over \partial
\tau}$ defines the light cone Hamiltonian of QCD.  The light-cone momenta
$P^+ = P^0 + P^z$ and $P_\perp$ are
kinematical and commute with $P^-$.
It is very useful to define the
invariant operator
$ H^{QCD}_{LC} = P^+ P^- - {\vec P_\perp}^2$ since its set of
eigenvalues
${\cal M}^2_n$ in the color-singlet sector of QCD enumerates the
bound state and continuum (scattering state) hadronic mass spectrum.
The solutions of the light-cone eigenvalue problem
$ H^{QCD}_{LC}
\ket{\psi_p} = M^2_p
\ket{\psi_p}$ are independent of $P^+$ and ${\vec P_\perp}$;  thus given
the eigensolution Fock projections $ \langle n; x_i, {\vec k_{\perp i}},
\lambda_i |\psi_p \rangle = \psi_n(x_i, {\vec k_{\perp i}},
\lambda_i) ,$ the wavefunction of the proton is determined in any
frame.\cite{Lepage:1980fj}
For example, the
projection of the proton eigensolution $\ket{\psi_p}$ on the
color-singlet
$B = 1$, $Q = 1$ eigenstates $\{\ket{n} \}$
of the free Hamiltonian $ H^{QCD}_{LC}(g = 0)$ gives the
light-cone Fock expansion:
$$ \left\vert \psi_p(P^+, {\vec P_\perp} )\right> = \sum_n \psi_n(x_i,
{\vec k_{\perp i}},
\lambda_i) \left\vert n;
x_i P^+, x_i {\vec P_\perp} + {\vec k_{\perp i}}, \lambda_i\right >. $$
The light-cone momentum fractions
$x_i = k^+_i/P^+$ with $\sum^n_{i=1} x_i = 1$ and ${\vec k_{\perp i}}$ with
$\sum^n_{i=1} {\vec k_{\perp i}} = {\vec 0_\perp}$ represent the
relative momentum
coordinates of the QCD constituents independent of the total momentum of
the state.  The actual physical transverse momenta are
${\vec p_{\perp i}} = x_i {\vec P_\perp} + {\vec k_{\perp i}}.$ The
$\lambda_i$ label the light-cone spin $S^z$ projections of the quarks and
gluons along the quantization $z$ direction.  The physical gluon
polarization vectors
$\epsilon^\mu(k,\ \lambda = \pm 1)$ are specified in light-cone
gauge by the conditions $k \cdot \epsilon = 0,\ \eta \cdot \epsilon =
\epsilon^+ = 0.$ Although light-cone gauge $A^+ = 0$ is traditionally
used in light-cone quantization, Srivastava and I have shown that one can
also effectively quantize QCD in the covariant Feynman
gauge.\cite{Srivastava:2000gi}

Each light-cone Fock wavefunction satisfies conservation of the
$z$ projection of angular momentum:
$
J^z = \sum^n_{i=1} S^z_i + \sum^{n-1}_{j=1} l^z_j \ .
$
The sum over $s^z_i$
represents the contribution of the intrinsic spins of the $n$ Fock state
constituents.  The sum over orbital angular momenta
$l^z_j = -{\mathrm i} (k^1_j\frac{\partial}{\partial k^2_j}
-k^2_j\frac{\partial}{\partial k^1_j})$ derives from
the $n-1$ relative momenta.  This excludes the contribution to the
orbital angular momentum due to the motion of the center of mass, which
is not an intrinsic property of the hadron.\cite{Brodsky:2000ii}

Recently Hwang, Ma,  Schmidt, and I have shown that
the light-cone wavefunctions generated by the radiative corrections to
the electron in QED provides an ideal system for understanding
the spin and angular momentum decomposition of relativistic
systems.\cite{Brodsky:2000ii} The model
is patterned after the quantum structure which
occurs in the one-loop Schwinger
${\alpha / 2 \pi} $ correction to the electron magnetic
moment.\cite{Brodsky:1980zm} In effect, we can represent a spin-\half ~
system as a composite of a spin-\half ~ fermion and spin-one vector boson
with arbitrary masses.  A similar model has recently been used to
illustrate the matrix elements and evolution of light-cone helicity and
orbital angular momentum operators.\cite{Harindranath:1999ve} This
representation of a composite system is particularly useful because it
is based on two constituents but yet is totally relativistic.  We can
then explicitly compute the form factors
$F_1(q^2)$ and $F_2(q^2)$ of the electromagnetic current, and the
various contributions to the form factors
$A(q^2)$ and $B(q^2)$ of the energy-momentum tensor.  The
anomalous moment coupling $B(0)$ to a graviton is shown to vanish for
any composite system.  This remarkable
result, first derived by Okun and Kobzarev,
\cite{Okun,Ji:1996kb,Ji:1997ek,Ji:1997nm,Teryaev:1999su} is shown to
follow directly from the Lorentz
boost properties of the light-cone Fock
representation.\cite{Brodsky:2000ii}

The light-cone Fock state wavefunctions of an electron can be
systematically evaluated in QED
perturbation theory~\cite{Lepage:1980fj,Brodsky:1980zm}:
The two-particle Fock state for an electron with $J^z =
+ {1\over 2}$ has four possible spin combinations:
\begin{equation}
\left
\{ \begin{array}{l}
\psi^{\uparrow}_{+\frac{1}{2}\, +1} (x,{\vec k}_{\perp})=-{\sqrt{2}}
\frac{(-k^1+{\mathrm i} k^2)}{x(1-x)}\,
\varphi \ , [\ell^z = -1]\\
\psi^{\uparrow}_{+\frac{1}{2}\, -1} (x,{\vec k}_{\perp})=-{\sqrt{2}}
\frac{(+k^1+{\mathrm i} k^2)}{1-x }\,
\varphi \ ,[\ell^z = +1] \\
\psi^{\uparrow}_{-\frac{1}{2}\, +1} (x,{\vec k}_{\perp})=-{\sqrt{2}}
(M-{m\over x})\,
\varphi \ ,[\ell^z = 0] \\
\psi^{\uparrow}_{-\frac{1}{2}\, -1} (x,{\vec k}_{\perp})=0\ ,
\end{array}
\right.
\label{vsn2}
\end{equation}
where
\begin{equation}
\varphi=\varphi (x,{\vec k}_{\perp})=\frac{ e/\sqrt{1-x}}{M^2-({\vec
k}_{\perp}^2+m^2)/x-({\vec k}_{\perp}^2+\lambda^2)/(1-x)}\ .
\label{wfdenom}
\end{equation}
Each configuration satisfies
the spin sum rule: $J^z=S^z_{\rm f}+s^z_{\rm b} + l^z = +{1\over 2}$.
The sign of the helicity of the
electron is retained by the leading photon at $x_\gamma = 1- x \to 1$.
Note that in the non-relativistic limit, the transverse motion of the
constituents can be neglected, and we have only the
$\left|+\frac{1}{2}\right> \to
\left|-\frac{1}{2}\, +1\right>$ configuration which is the
non-relativistic quantum state for the spin-half system composed of
a fermion and a spin-1 boson constituents.  The fermion
constituent has spin projection in the opposite
direction to the spin $J^z$ of the whole system.
However, for ultra-relativistic binding in which
the transversal motions of the constituents are large compared to the
fermion masses,  the
$\left|+\frac{1}{2}\right> \to \left|+\frac{1}{2}\, +1\right>$
and
$\left|+\frac{1}{2}\right> \to \left|+\frac{1}{2}\, -1\right>$
configurations dominate
over the $\left|+\frac{1}{2}\right> \to \left|-\frac{1}{2}\, +1\right>$
configuration.  In this case
the fermion constituent has
spin projection parallel to $J^z$.

The spin
structure of perturbative theory provides a
template for the numerator structure of the light-cone wavefunctions
even for composite systems since the equations which couple
different Fock components mimic the perturbative form.
The structure of the electron's Fock state in perturbative QED shows that
it is natural to have a negative contribution from relative orbital
angular momentum which balances the $S_z$ of its photon constituents.
We can thus expect a large orbital contribution to the proton's
$J_z$ since gluons carry roughly half of the proton's
momentum, thus providing insight into the ``spin crisis" in QCD.

The light-cone Fock representation of current matrix elements has a number
of simplifying properties.  Matrix elements of space-like local operators
for the coupling of photons, gravitons and the deep inelastic
structure functions can all be expressed as overlaps of light-cone
wavefunctions with the same number of Fock constituents.  This is possible
since one can choose the special frame $q^+ = 0$
\cite{Drell:1970km,West:1970av} for space-like momentum transfer and
take matrix elements of ``plus" components of currents such as $J^+$ and
$T^{++}$.  Since the physical vacuum in light-cone quantization coincides
with the perturbative vacuum, no contributions to matrix elements from
vacuum fluctuations occur.\cite{Brodsky:1998de}   Exclusive
semi-leptonic
$B$-decay amplitudes involving timelike currents such as $B\rightarrow A
\ell
\bar{\nu}$ can also be evaluated exactly.\cite{Brodsky:1999hn,Ji:1999gt}
In this case, the timelike decay matrix elements require the
computation of both the diagonal matrix element $n \rightarrow n$ where
parton number is conserved and the off-diagonal $n+1\rightarrow n-1$
convolution such that the current operator annihilates a $q{\bar{q'}}$
pair in the initial $B$ wavefunction.  This term is a consequence of the
fact that the time-like decay $q^2 = (p_\ell + p_{\bar{\nu}} )^2 > 0$
requires a positive light-cone momentum fraction $q^+ > 0$.  A similar
result holds for the light-cone wavefunction representation of the deeply
virtual Compton amplitude.\cite{DVCSHwang}

Given the light-cone wavefunctions, one can compute the
moments of the helicity and
transversity distributions measurable in polarized deep inelastic
experiments.\cite{Lepage:1980fj}   For example,
the polarized quark distributions at resolution $\Lambda$ correspond to
\begin{eqnarray}
q_{\lambda_q/\Lambda_p}(x, \Lambda)
&=& \sum_{n,q_a}
\int\prod^n_{j=1} dx_j d^2 k_{\perp j}\sum_{\lambda_i}
\vert \psi^{(\Lambda)}_{n/H}(x_i,\vec k_{\perp i},\lambda_i)\vert^2
\\
&& \times \delta\left(1- \sum^n_i x_i\right) \delta^{(2)}
\left(\sum^n_i \vec k_{\perp i}\right)
\delta(x - x_q) \delta_{\lambda_a, \lambda_q}
\Theta(\Lambda^2 - {\cal M}^2_n)\ , \nonumber
\end{eqnarray}
where the sum is over all quarks $q_a$ which match the quantum
numbers, light-cone momentum fraction $x,$ and helicity of the struck
quark.  Similarly, the distribution of spectator
particles in the final state which could be measured in the proton
fragmentation region in deep inelastic scattering at an electron-proton
collider are in principle encoded in the light-cone wavefunctions.

The key non-perturbative input for exclusive
processes is the gauge and frame independent hadron distribution
amplitude \cite{Lepage:1979zb,Lepage:1980fj} defined as the integral of
the valence (lowest particle number) Fock wavefunction;
\eg\ for the pion
\begin{equation}
\phi_\pi (x_i,\Lambda) \equiv \int d^2k_\perp\, \psi^{(\Lambda)}_{q\bar
q/\pi} (x_i, \vec k_{\perp i},\lambda)
\label{eq:f1}
\end{equation}
where the global cutoff $\Lambda$ is identified with the resolution $Q$.
The distribution amplitude controls leading-twist exclusive amplitudes
at high momentum transfer, and it can be related to the gauge-invariant
Bethe-Salpeter wavefunction at equal light-cone time.  The
logarithmic evolution of hadron distribution amplitudes
$\phi_H (x_i,Q)$ can be derived from the perturbatively-computable tail
of the valence light-cone wavefunction in the high transverse momentum
regime.\cite{Lepage:1979zb,Lepage:1980fj} The conformal basis for the
evolution of the three-quark distribution amplitudes for
the baryons~\cite{Lepage:1979za} has recently been obtained by V. Braun
\etal.\cite{Braun:1999te}

The main features of the heavy sea quark-pair contributions of the
higher particle number Fock state states of light hadrons can be
derived from perturbative QCD, One can identify two contributions to the
heavy quark sea, the ``extrinsic'' contributions which correspond to
ordinary gluon splitting, and the ``intrinsic" sea which is
multi-connected via gluons to the valence quarks.  The leading $1/m_Q^2$
contributions to the intrinsic sea of the proton in the heavy quark
expansion are proton matrix elements of the operator~\cite{Franz:2000ee}
$\eta^\mu \eta^\nu G_{\alpha \mu} G_{\beta \nu} G^{\alpha \beta}$ which
in light-cone gauge $\eta^\mu A_\mu= A^+= 0$ corresponds to three or four
gluon exchange between the heavy-quark loop and the proton constituents
in the forward virtual Compton amplitude.  The intrinsic sea is thus
sensitive to the hadronic bound-state structure.\cite{Brodsky:1981se} The
maximal contribution of the intrinsic heavy quark occurs at $x_Q \simeq
{m_{\perp Q}/ \sum_i m_\perp}$ where
$m_\perp = \sqrt{m^2+k^2_\perp}$;
\ie\ at large $x_Q$, since this minimizes the invariant mass $\M^2_n$.
The
measurements of the charm structure function by the EMC experiment are
consistent with intrinsic charm at large $x$ in the nucleon with a
probability of order $0.6 \pm 0.3 \% $.\cite{Harris:1996jx} which is
consistent with recent estimates based on instanton
fluctuations.\cite{Franz:2000ee} Similarly, one can distinguish
intrinsic gluons which are associated with multi-quark interactions and
extrinsic gluon contributions associated with quark
substructure.\cite{Brodsky:1990db} One can also use this framework to
isolate the physics of the anomaly contribution to the Ellis-Jaffe sum
rule.\cite{Bass:1998rn} Thus neither gluons nor sea quarks are solely
generated by DGLAP evolution, and one cannot define a resolution scale
$Q_0$ where the sea or gluon degrees of freedom can be neglected.

It is usually assumed that a heavy quarkonium state such as the
$J/\psi$ always decays to light hadrons via the annihilation of its heavy quark
constituents to gluons.  However, as Karliner and I \cite{Brodsky:1997fj}
have shown, the transition $J/\psi \to \rho
\pi$ can also occur by the rearrangement of the $c \bar c$ from the $J/\psi$
into the $\ket{ q \bar q c \bar c}$ intrinsic charm Fock state of the $\rho$ or
$\pi$.  On the other hand, the overlap rearrangement integral in the
decay $\psi^\prime \to \rho \pi$ will be suppressed since the intrinsic
charm Fock state radial wavefunction of the light hadrons will evidently
not have nodes in its radial wavefunction.  This observation provides
a natural explanation of the long-standing puzzle~\cite{Brodsky:1987bb}
why the $J/\psi$ decays prominently to two-body pseudoscalar-vector final
states, breaking hadron helicity
conservation,~\cite{Brodsky:1981kj} whereas the
$\psi^\prime$ does not.

\section{ Applications of Light-Cone Factorization to Hard QCD Processes}

The light-cone
formalism provides a physical factorization scheme which conveniently
separates and factorizes soft non-perturbative physics from hard
perturbative dynamics in both exclusive and
inclusive reactions.\cite{Lepage:1980fj,Lepage:1979zb} In hard inclusive
reactions all intermediate states are divided according to $\M^2_n <
\Lambda^2 $ and
$\M^2_n >
\Lambda^2 $ domains.  The lower mass regime is associated with the quark
and gluon distributions defined from the absolute squares of the LC
wavefunctions in the light cone factorization scheme.  In the high
invariant mass regime, intrinsic transverse momenta can be ignored, so
that the structure of the process at leading power has the form of hard
scattering on collinear quark and gluon constituents, as in the parton
model.  The attachment of gluons from the LC wavefunction to a propagator
in a hard subprocess is power-law suppressed in LC gauge, so that the
minimal quark-gluon particle-number subprocesses dominate.

There are many applications of this formalism:
{\it Exclusive Processes and Heavy Hadron Decays.} At high transverse
momentum an exclusive amplitudes factorize into a convolution of a hard
quark-gluon subprocess amplitudes $T_H$ with the hadron distribution
amplitudes
$\phi(x_i,\Lambda)$.  {\it Color Transparency.} Each Fock state interacts
distinctly; \eg\ Fock states with small particle number and small impact
separation have small color dipole moments and can traverse a nucleus
with minimal interactions.  This is the basis for the predictions for
color transparency \cite{BM} in hard quasi-exclusive reactions.  {\it
Diffractive vector meson photoproduction.} The light-cone Fock
wavefunction representation of hadronic amplitudes allows a simple
eikonal analysis of diffractive high energy processes, such as
$\gamma^*(Q^2) p \to \rho p$, in terms of the virtual photon and the vector
meson Fock state light-cone wavefunctions convoluted with the $g p \to g p$
near-forward matrix element.\cite{Brodsky:1994kf}
{\it Regge behavior of structure functions.} The light-cone wavefunctions
$\psi_{n/H}$ of a hadron are not independent of each other, but rather are
coupled via the QCD equations of motion.  The constraint of finite
``mechanical'' kinetic energy implies ``ladder relations"
which interrelate the light-cone wavefunctions of states differing by one
or two gluons.\cite{ABD} This in turn implies BFKL
Regge behavior of the polarized and unpolarized structure functions at $x
\rightarrow 0$.\cite{Mueller}
{\it Structure functions at large $x_{bj}$.} The behavior of structure functions
at $x \rightarrow 1$ is a highly off-shell light-cone
wavefunction configuration leading to quark-counting and
helicity-retention rules for the power-law behavior of the polarized and
unpolarized quark and gluon distributions in the endpoint domain.  The
effective starting point for the PQCD evolution of the structure
functions increases as $x \to 1$.  Thus evolution
is quenched at $ x \to 1$.\cite{Lepage:1980fj,BrodskyLepage,Dmuller}
{\it Hidden Color.}
The deuteron form factor at high $Q^2$ is sensitive to wavefunction
configurations where all six quarks overlap within an impact
separation $b_{\perp i} < {\cal O} (1/Q).$ The dominant
color configuration at large distances corresponds to the usual
proton-neutron bound state.  However, at small impact space
separation, all five Fock color-singlet components eventually
acquire equal weight, \ie, the deuteron wavefunction evolves to
80\%\ ``hidden color.'' The relatively large normalization of the
deuteron form factor observed at large $Q^2$ hints at sizable
hidden-color contributions.\cite{Farrar:1991qi} Hidden color components
can also play a predominant role in the reaction $\gamma d \to J/\psi p n$
at threshold if it is dominated by the multi-fusion process $\gamma g g
\to J/\psi$.

\section{Self-Resolved Diffractive Reactions and Light Cone Wavefunctions}

Diffractive multi-jet production in heavy
nuclei provides a novel way to measure the shape of the LC Fock
state wavefunctions and test color transparency.  For example, consider the
reaction \cite{Bertsch,MillerFrankfurtStrikman,Frankfurt:1999tq}
$\pi A \rightarrow {\rm Jet}_1 + {\rm Jet}_2 + A^\prime$
at high energy where the nucleus $A^\prime$ is left intact in its ground
state.  The transverse momenta of the jets balance so that
$
\vec k_{\perp i} + \vec k_{\perp 2} = \vec q_\perp < {R^{-1}}_A \ .
$
The light-cone longitudinal momentum fractions also need to add to
$x_1+x_2 \sim 1$ so that $\Delta p_L < R^{-1}_A$.  The process can
then occur coherently in the nucleus.  Because of color transparency, the
valence wavefunction of the pion with small impact separation, will
penetrate the nucleus with minimal interactions, diffracting into jet
pairs.\cite{Bertsch} The $x_1=x$, $x_2=1-x$ dependence of
the di-jet distributions will thus reflect the shape of the pion
valence light-cone wavefunction in $x$; similarly, the
$\vec k_{\perp 1}- \vec k_{\perp 2}$ relative transverse momenta of the
jets gives key information on the derivative of the underlying shape
of the valence pion
wavefunction.\cite{MillerFrankfurtStrikman,Frankfurt:1999tq,BHDP} The
diffractive nuclear amplitude extrapolated to
$t = 0$ should be linear in nuclear number $A$ if color transparency is
correct.  The integrated diffractive rate should then scale as $A^2/R^2_A
\sim A^{4/3}$.  Preliminary results on a diffractive dissociation
experiment of this type E791 at Fermilab using 500 GeV incident pions on
nuclear targets.\cite{Ashery:1999nq} appear to be consistent with color
transparency.\cite{Ashery:1999nq} The momentum fraction distribution of
the jets is consistent with a valence light-cone wavefunction of the pion
consistent with the shape of the asymptotic distribution amplitude,
$\phi^{\rm asympt}_\pi (x) =
\sqrt 3 f_\pi x(1-x)$.  Data from
CLEO \cite{Gronberg:1998fj} for the
$\gamma
\gamma^* \rightarrow \pi^0$ transition form factor also favor a form for
the pion distribution amplitude close to the asymptotic solution
\cite{Lepage:1979zb,Lepage:1980fj} to the perturbative QCD evolution
equation.

The diffractive dissociation of a hadron or nucleus can also occur via
the Coulomb dissociation of a beam particle on an electron beam (\eg\ at
HERA or eRHIC) or on the strong Coulomb field of a heavy nucleus (\eg\
at RHIC or nuclear collisions at the LHC).\cite{BHDP} The amplitude for
Coulomb exchange at small momentum transfer is proportional to the first
derivative $\sum_i e_i {\partial \over \vec k_{T i}} \psi$ of the
light-cone wavefunction, summed over the charged constituents.  The Coulomb
exchange reactions fall off less fast at high transverse momentum compared
to pomeron exchange reactions since the light-cone wavefunction is
effective differentiated twice in two-gluon exchange reactions.

It will also be interesting to study diffractive tri-jet production
using proton beams
$ p A \rightarrow {\rm Jet}_1 + {\rm Jet}_2 + {\rm Jet}_3 + A^\prime $ to
determine the fundamental shape of the 3-quark structure of the valence
light-cone wavefunction of the nucleon at small transverse
separation.\cite{MillerFrankfurtStrikman}
For example, consider the Coulomb dissociation of a high energy proton at
HERA.  The proton can dissociate into three jets corresponding to the
three-quark structure of the valence light-cone wavefunction.  We can
demand that the produced hadrons all fall outside an opening angle $\theta$
in the proton's fragmentation region.
Effectively all of the light-cone momentum
$\sum_j x_j \simeq 1$ of the proton's fragments will thus be
produced outside an ``exclusion cone".  This
then limits the invariant mass of the contributing Fock state ${\cal
M}^2_n >
\Lambda^2 = P^{+2} \sin^2\theta/4$ from below, so that perturbative QCD
counting rules can predict the fall-off in the jet system invariant mass
$\cal M$.  At large invariant mass one expects the three-quark valence
Fock state of the proton to dominate.  The segmentation of the forward
detector in azimuthal angle $\phi$ can be used to identify structure and
correlations associated with the three-quark light-cone
wavefunction.\cite{BHDP}
An interesting possibility is
that the distribution amplitude of the
$\Delta(1232)$ for $J_z = 1/2, 3/2$ is close to the asymptotic form $x_1
x_2 x_3$,  but that the proton distribution amplitude is more complex.
This ansatz can also be motivated by assuming a quark-diquark structure
of the baryon wavefunctions.  The differences in shapes of the
distribution amplitudes could explain why the $p
\to\Delta$ transition form factor appears to fall faster at large $Q^2$
than the elastic $p \to p$ and the other $p \to N^*$ transition form
factors.\cite{Stoler:1999nj} One can use also measure the dijet
structure of real and virtual photons beams
$ \gamma^* A \rightarrow {\rm Jet}_1 + {\rm Jet}_2 + A^\prime $ to
measure the shape of the light-cone wavefunction for
transversely-polarized and longitudinally-polarized virtual photons.  Such
experiments will open up a direct window on the amplitude
structure of hadrons at short distances.
The light-cone formalism is also applicable to the
description of nuclei in terms of their nucleonic and mesonic
degrees of freedom.\cite{Miller:1999mi,Miller:2000ta}
Self-resolving diffractive jet reactions
in high energy electron-nucleus collisions and hadron-nucleus collisions
at moderate momentum transfers can thus be used to resolve the light-cone
wavefunctions of nuclei.

\section{Non-Perturbative Solutions of Light-Cone Quantized QCD}

Is there any hope of computing light-cone wavefunctions from
first principles?  In the discretized light-cone quantization
method (DLCQ),\cite{Pauli:1985ps}
periodic boundary conditions are introduced in
$b_\perp$ and $x^-$ so that the momenta
$k_{\perp i} = n_\perp \pi/ L_\perp$ and $x^+_i = n_i/K$ are
discrete.  A global cutoff in invariant mass of the partons in the Fock
expansion is also introduced.
Solving the quantum field theory then reduces to
the problem of diagonalizing the finite-dimensional hermitian matrix
$H_{LC}$ on a finite discrete Fock basis.  The DLCQ method has now become
a standard tool for solving both the spectrum and light-cone wavefunctions
of one-space one-time theories -- virtually any
$1+1$ quantum field theory, including ``reduced QCD" (which has both quark and
gluonic degrees of freedom) can be completely solved using
DLCQ.\cite{Dalley:1993yy,AD} Hiller, McCartor, and I
\cite{Brodsky:1998hs,Brodsky:1999xj} have recently shown that the use of
covariant Pauli-Villars regularization with DLCQ allows one to obtain the
spectrum and light-cone wavefunctions of simplified theories in physical
space-time dimensions, such as (3+1) Yukawa theory.  Dalley \etal\ have
also showed how one can use DLCQ with a transverse lattice to solve
gluonic $ 3+1$ QCD.\cite{Dalley:1999ii}
The spectrum obtained for gluonium states is in remarkable
agreement with lattice gauge theory results, but with a huge reduction of
numerical effort.  Hiller and I \cite{Hiller:1999cv} have shown how one
can use DLCQ to compute the electron magnetic moment in QED without
resort to perturbation theory.  One can also formulate DLCQ so that
supersymmetry is exactly preserved in the discrete approximation, thus
combining the power of DLCQ with the beauty of
supersymmetry.\cite{Lunin:1999ib,Haney:1999tk} The ``SDLCQ" method has
been applied to several interesting supersymmetric theories, to the
analysis of zero modes, vacuum degeneracy, massless states, mass gaps,
and theories in higher dimensions, and even tests of the Maldacena
conjecture.\cite{Antonuccio:1999ia} Broken supersymmetry is interesting
in DLCQ, since it may serve as a method for regulating non-Abelian
theories.\cite{Brodsky:1999xj}
Another
remarkable advantage of light-cone quantization is that the vacuum state
$\ket{0}$ of the full QCD Hamiltonian coincides with the free vacuum.
For example, as discussed by Bassetto,\cite{Bassetto:1999tm}
the computation of the spectrum
of $QCD(1+1)$ in equal time quantization requires constructing the full
spectrum of non perturbative contributions (instantons).  However,
light-cone methods with infrared regularization give the correct result
without any need for vacuum-related contributions.  The role
of instantons and such phenomena in light-cone quantized
$QCD(3+1)$ is presumably more complicated and may reside in zero
modes;\cite{Yamawaki:1998cy}
\eg, zero modes are evidently necessary to represent theories with
spontaneous symmetry breaking.\cite{Pinsky:1994yi}

Even without full non-perturbative solutions of QCD, one can envision a
program to construct the light-cone wavefunctions using measured moments
constraints from QCD sum rules, lattice gauge theory,  hard
exclusive and inclusive processes.  One is guided by theoretical
constraints from perturbation theory which dictates the asymptotic form
of the wavefunctions at large invariant mass,
$x \to 1$, and high
$k_\perp$.\cite{Lepage:1980fj,Hoyer:1990pa}   One can also use constraints
from ladder relations which connect Fock states of
different particle number; perturbatively-motivated numerator spin
structures; guidance from toy models
such as ``reduced"
$QCD(1+1)$~\cite{AD}; and the correspondence to Abelian theory
for
$N_C\to 0$~\cite{Brodsky:1998jk} and the many-body
Schr\"odinger theory in the nonrelativistic domain.

In this talk I have
discussed how the universal, process-independent and
frame-independent light-cone Fock-state wavefunctions
encode the properties of a hadron in terms of its fundamental quark and
gluon degrees of freedom.  Given the proton's light-cone wavefunctions,
one can compute not only the moments of the quark and gluon distributions
measured in deep inelastic lepton-proton scattering, but also the
multi-parton correlations which control the distribution of particles in
the proton fragmentation region and dynamical higher twist effects.
Light-cone wavefunctions also provide a systematic framework for
evaluating exclusive hadronic matrix elements, including time-like heavy
hadron decay amplitudes and form factors.  The formalism also provides a
physical factorization scheme for separating hard and soft contributions
in both exclusive and inclusive hard processes.  A new type of jet
production reaction, ``self-resolving diffractive interactions" can
provide direct information on the light-cone wavefunctions of hadrons in
terms of their QCD degrees of freedom, as well as the composition of
nuclei in terms of their nucleon and mesonic degrees of freedom.

\section*{Acknowledgments}
Work supported by the Department of Energy
under contract number DE-AC03-76SF00515.
I wish to thank Tony Kennedy, Andreas Schreiber, and Tony Thomas
for their kind hospitality at the University of Adelaide.

\end{document}